%% file: paper-aiwild.tex
\documentclass{article}

\usepackage{microtype}
\usepackage{graphicx}
\usepackage{subcaption}
\usepackage{booktabs} 
\usepackage{array}
\usepackage{float}
\usepackage{xcolor}
\usepackage{tabularx}
\usepackage{longtable}
\definecolor{viewColor}{HTML}{2F7D5C}
\definecolor{baseColor}{HTML}{B04545}
\definecolor{changesColor}{HTML}{3D5A94}

\PassOptionsToPackage{hyphens}{url}
\usepackage{hyperref}
\usepackage{xurl}
\makeatletter
\g@addto@macro\UrlBreaks{\do\-\do\.\do\=\do\&\do\?\do\_\do\/}
\makeatother
\usepackage{algorithm}
\usepackage{algorithmic}

\usepackage[accepted]{icml2026}

\usepackage{amsmath}
\usepackage{amssymb}
\usepackage{mathtools}
\usepackage{amsthm}
\usepackage{enumitem}

\usepackage[capitalize,noabbrev]{cleveref}

\theoremstyle{plain}

\theoremstyle{definition}

\theoremstyle{remark}

\usepackage[textsize=tiny]{todonotes}

\icmltitlerunning{Copy-on-Write Scoring: Application-Specific Agent Evaluations}

\begin{document}

\twocolumn[
  \icmltitle{Copy-on-Write Scoring: Application-Specific Agent Evaluations}




  \begin{icmlauthorlist}
    \icmlauthor{Joanna Roy}{trail}
    \icmlauthor{Sven Hölzel}{trail}
  \end{icmlauthorlist}

  \icmlaffiliation{trail}{trail-ml, Munich, Germany}

  \icmlcorrespondingauthor{Joanna Roy}{joanna@trail-ml.com}
  \icmlcorrespondingauthor{Sven Hölzel}{sven@trail-ml.com}

  \icmlkeywords{LLM agents, evaluation, Copy-on-Write, benchmarking, software systems}

  \vskip 0.3in
]



\printAffiliationsAndNotice{}  

\begin{abstract}
  Trustworthy deployment of LLM-based agents in software systems requires evaluating how they perform on application-specific workflows, with enough granularity to localize where they succeed and fail. Yet existing agent evaluation mechanisms are limited: benchmarks have low construct validity for application-specific workflows and environments, and replica evaluation environments are expensive and prone to drift. We propose \textbf{Copy-on-Write (CoW) Scoring}\footnote{Python library: \href{https://github.com/trail-ml/agent-cow-python}{\texttt{agent-cow}}}, a framework that evaluates agent operations directly within application environments using a PostgreSQL-level Copy-on-Write mechanism to isolate agent writes. CoW Scoring produces session- and operation-level scores that highlight where agents' database write operations succeed and fail in a given application environment, enabling inexpensive evaluation and iteration on agent harnesses and tool surfaces. We demonstrate the framework on Plane, an open-source project-management platform, where analysis surfaced specific issues in the tool surface, and corresponding fixes produced measurable improvements on affected models.  
\end{abstract}

\section{Introduction}

Large Language Model (LLM)-based agents are increasingly deployed in production software systems, from enterprise tooling (e.g., ServiceNow, Salesforce, Notion) to MCP integrations across a wide range of consumer applications.\footnote{Gartner (2025); BCG (2025); Deloitte (2026)} Many of these applications are relied upon by a wide range of organizations and end-users, underpinning core business operations across large segments of the economy. Agent unreliability and errors --- such as accidental deletion or overwriting of data, misuse of tooling, or incorrect changes to critical records --- could have substantial ripple effects, yet are difficult to identify and prevent in practice, particularly when agent writes are interleaved with the prior database state. 

Evaluating agents in software contexts remains an underdeveloped area. Benchmarks are the most common means of assessing performance, but strong benchmark results do not guarantee strong performance in real systems or on real-world tasks \cite{mohammadiEvaluationBenchmarkingLLM2025, rajiAI2021,liaoRethinking2025, dengInvestigating2023}. Existing benchmarks tend to have low \emph{construct validity} \cite{beanMeasuring2025} for the environments in which software agents are deployed.

Some benchmarks have attempted to close this gap by situating agents in replicas of common applications \cite{drouinWorkArena2024, zhouWebArena2024, xuTheAgentCompany2024} or in simulated domains with API access \cite{yao$t$bench2024}. While these partly address the construct validity gap, two main limitations remain: (a) replicas and simulations are time- and resource-expensive to produce, and can quickly drift from the live state of the application; and (b) evaluating performance on a \emph{specific} software is still not necessarily predictive of performance in others --- for example, one with different tooling, system prompts, and workflows. Since no single benchmark can predict agent performance for arbitrary applications, teams need complementary evaluation methods that work directly in their application environment, on representative workflows.

We propose \emph{\textbf{Copy-on-Write (CoW) Scoring}}, a framework to develop and safely conduct use-case-specific agent evaluations in software applications. The CoW mechanism allows agents to operate directly within the application environment, avoiding the cost and drift of replicas while keeping sessions isolated and setup easy to reuse across runs. This work assumes a PostgreSQL database, although the CoW pattern could similarly be extended to other data stores, which would enable scoring a broader range of agent operations. The scoring framework provides session- and operation-level scores that surface where agent writes succeed and fail when interacting with a given tool surface. 

We demonstrate an evaluation loop on Plane,\footnote{\href{https://plane.so/}{plane.so}} an open-source project-management platform, where scoring surfaces multiple failure modes specific to the deployed agents and tool surface, and a corresponding fix produces measurable improvement on the affected models. The same score-diagnose-fix procedure applies to any application using CoW Scoring, enabling low-cost iteration on the model, prompt, retrieval, or system prompt in a given deployment. 

\section{Copy-on-Write (CoW) Scoring}
\label{sec:cow-scoring}

This section describes the CoW \emph{mechanism}, which isolates agent database changes, and the CoW \emph{Scoring framework}, which compares a ground-truth (GT) execution of a given workflow with a corresponding agent execution at the session and operation level. Both components are implemented in the open-source \texttt{agent-cow} library.

\subsection{CoW Mechanism}
\label{sec:cow}

Copy-on-Write (CoW) is a resource-management technique that lets processes share underlying data without creating copies upfront. Rather, reads go to the shared data and writes trigger creation of a copy such that only the writing process observes the change \cite{silberschatzOperatingSystemConcepts}. Applied to agent operations on databases, CoW prevents agents from writing directly to the underlying production data. 

To enable CoW on a database, each original table is split into a \textbf{base} table (contains the underlying production data), \textbf{changes} table (contains agent writes), and corresponding \textbf{view} (defined by a query merging base and changes table data). All application operations are sent to the view, since it takes the name of the original table. View triggers redirect CoW write operations to the changes table, with associated CoW metadata columns (\texttt{session\_id}, \texttt{operation\_id}, \texttt{\_cow\_updated\_at}, \texttt{\_cow\_deleted}) appended. Actions from a given session can then be queried using the respective \texttt{session\_id}. The components are visualized in Figure~\ref{fig:cow}, and more details are described in Appendix~\ref{app:cow}.

\begin{figure}[t]
  \centering
  \includegraphics[width=\columnwidth]{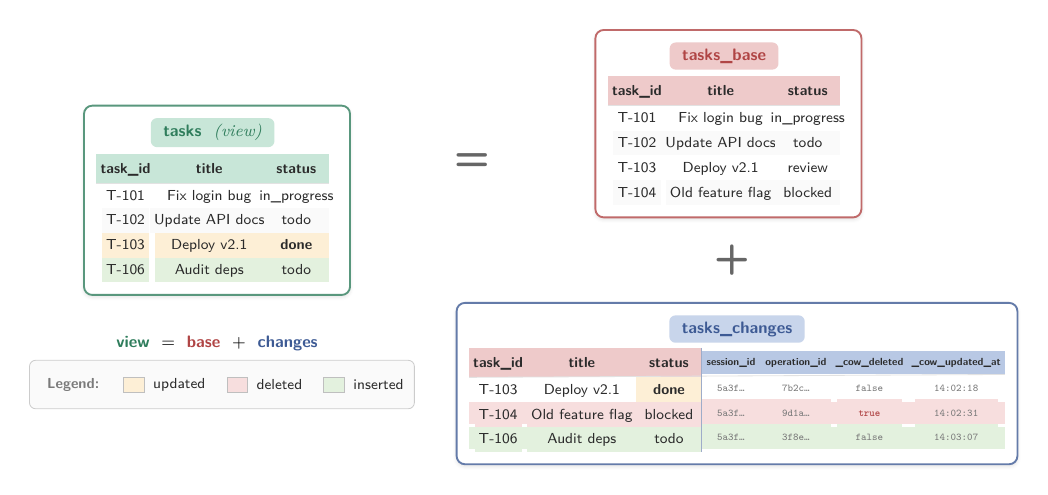}
  \caption{
    \textbf{Copy-on-Write (CoW) mechanism in agent evaluation.}
    The agent reads and writes through a 
    \textcolor{viewColor}{\textbf{view}}, which merges the
    \textcolor{baseColor}{\textbf{base table}} (production data) with the
    \textcolor{changesColor}{\textbf{changes table}} (per-session using the operation/session metadata).
    Agent writes are intercepted and written to the changes table, leaving base data untouched.
  }
  \label{fig:cow}
\end{figure}

The \texttt{agent-cow} library handles database setup --- creating base tables, changes tables, views, and deploying triggers (Algorithm~\ref{alg:cow-enable}, Appendix~\ref{app:cow}), and the application is responsible for passing \texttt{session\_id} and \texttt{operation\_id} with each agent operation. For Plane, CoW integration required $\sim$250 lines of code for CoW functionality, and an additional $\sim$540 lines specific to our recording infrastructure and portable to a testing harness (Appendix~\ref{app:cow-plane}). We have separately integrated CoW Scoring into a closed-source production application, with comparable effort.

\subsection{Scoring}
\label{sec:scoring}

Scoring takes two CoW sessions as input: a ground-truth (GT) session and an agent session. The GT session is recorded by a human performing the ideal sequence of actions for a given workflow with CoW enabled. The human also writes a prompt describing the workflow such that they would expect an agent to be able to reproduce their actions. The agent is then given this prompt and executes its own sequence of actions with CoW enabled.  

CoW Scoring happens along two dimensions and at two levels of granularity. The two dimensions are structural (which tables, rows, columns, and relationships were written) and content (what values were written). The two levels of granularity are session-level (evaluates the session as a whole) and operation-level (evaluates each operation). Users can supply their own comparators and overall scoring functions; Appendix~\ref{app:cow-scoring-lib} describes the library architecture.

\subsubsection{Session-level comparison}

For session-level scoring, we keep the most recent written row state for each primary key. Changed rows are classified as: \emph{matched} ($N_\mathrm{matched}$ --- rows present in both GT and agent sessions, under UUID mapping), \emph{missing} ($N_\mathrm{missing}$ --- GT rows the agent never produced), or \emph{extra} ($N_\mathrm{extra}$ --- agent rows with no GT counterpart).

\paragraph{Session-level structural score, $s^{\mathrm{struct}}$:} The fraction of \emph{matched} rows relative to the total number of rows changed.
\begin{equation}
    s^{\mathrm{struct}} = \frac{N_\mathrm{matched}}{N_\mathrm{matched} + N_\mathrm{missing} + N_\mathrm{extra}}
\end{equation}

\paragraph{Session-level content score, $s^{\mathrm{content}}$:} The mean per-field similarity over all matched rows.
\begin{equation}
    s^{\mathrm{content}} = \frac{1}{N_\mathrm{matched}} \sum_{(g, a) \in \text{matched}} \mathrm{sim}(g, a)
\end{equation}
where $\mathrm{sim}(g, a) \in [0, 1]$ is the similarity calculation per row between $g$ (GT) and $a$ (agent). Each column has a default similarity comparator according to its data type, but custom comparators can be provided as described in Appendix~\ref{app:customization}.

Notably, CoW Scoring compares final database states rather than action sequences, so the session-level scores are invariant to the path the agent took.

\subsubsection{Operation-level comparison}

While the session-level scores evaluate whether the agent reached the desired GT world state, the operation-level scores quantify how useful each operation was in getting there. We first sort operations in topological order, then calculate the following for each operation. 

\paragraph{Op-level structural utility, $u^{\mathrm{struct}}(o_i)$:} The change in the session-level structural score before and after $o_i$ is applied.

\begin{equation}
  u^{\mathrm{struct}}(o_i) = s^{\mathrm{struct}}(G_{\leq i}) - s^{\mathrm{struct}}(G_{< i})
\end{equation}

Where $G_{< i}$ denotes the world state prior to applying $o_i$, and $G_{\leq i}$ denotes the state after $o_i$ is applied. 

\paragraph{Op-level content utility $u^{\mathrm{content}}(o_i)$:} The mean per-field content similarity over operation rows.

\begin{equation}
  u^{\mathrm{content}}(o_i) = \frac{1}{|M_i|} \sum_{(g, a) \in M_i} \mathrm{sim}(g, a)
\end{equation}
where $M_i$ is the subset of matched rows whose agent-session write originated in $o_i$. 

\section{Methods}

Because the framework targets application-specific evaluation rather than a universal benchmark, our empirical validation will not compare against an external ground truth. The preliminary study below aims to show that (a) the CoW mechanism and scoring framework can be implemented in a PostgreSQL application, and (b) resulting scores surface useful information about agent behaviour, which can inform improvements to the tool surface in subsequent iterations.

\subsection{Preliminary Study: Application in Plane}

We apply CoW Scoring to \textbf{Plane}, an open-source project-management platform, as a concrete demonstration of the evaluation workflow. To produce a diverse set of ground-truth (GT) sessions without spending excessive author effort, Claude Opus 4.6 with access to the Plane codebase was prompted to execute realistic workflows against the seeded data, and each workflow manually reviewed and validated by the authors (details in Appendix~\ref{app:plane-integration}). We developed a testing harness, which embeds Plane's OpenAPI specification into a vector store, and exposes them to the agent through two tools: a \texttt{discover} tool for querying relevant endpoints, and an \texttt{execute} tool for making the corresponding requests. 

Since CoW sessions are isolated, each GT and agent session started with the same application state. For each GT session, an agent was given the associated prompt and access to the Plane API via the harness \texttt{execute} and \texttt{discover} tools, with a 50-operation limit per session. Five language models were evaluated twice per prompt: GPT-5, GPT-4.1, Gemini-3.1-Pro, Gemini-3.1-Flash-Lite, and Gemini-2.5-Pro. In total, $5 \times 2 \times 20 = 200$ agent sessions were scored against their corresponding GT sessions using CoW Scoring. A third trial per model was run after updating the tool surface (Section~\ref{sec:improvement-iteration}), for a total of 300 sessions.

\section{Preliminary Results}
\label{sec:results}

\subsection{Initial scores across models}

\begin{figure}[t]
  \centering
  \includegraphics[width=\columnwidth]{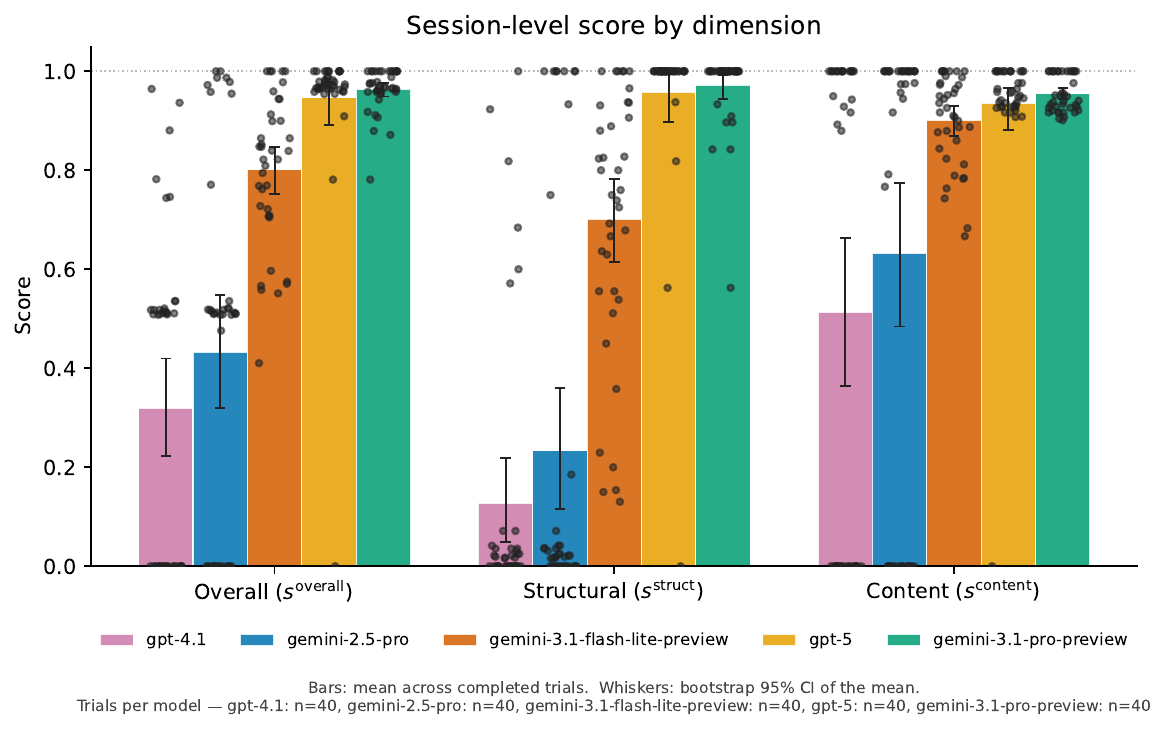}
  \caption{\textbf{Score breakdown} by dimension per model, for runs 1 and 2.}
  \label{fig:results-dimensions}
\end{figure}

Figure~\ref{fig:results-dimensions} shows CoW scores along each dimension, averaged between the two runs per (model, workflow) pair, with per-model summaries in Appendix~\ref{app:results-summary} (Tables~\ref{tab:results-summary} and~\ref{tab:worst-sessions}). These are early results from a single application and 20 workflows, with two trials per (model, workflow) pair, so broader claims would benefit from more samples.

Two initial observations serve as a sanity check on the scoring functions: within each model family, the ordering matches public benchmarks (Gemini-3.1-Pro $>$ Gemini-3.1-Flash-Lite, GPT-5 $>$ GPT-4.1), and per-model overall scores differ by at most 0.03 between trials, suggesting the means are representative rather than single-run artefacts.  

\subsection{Diagnosing failure modes}


\paragraph{Vocabulary mismatch.} Plane uses ``work item'' to describe tasks within projects, while the session prompts use the term ``issue'', which is also commonly used to describe these items (the database tables still use ``issues'', suggesting an artefact of a previous naming convention), so an agent that searched for ``issue'' would not immediately find relevant operations. \textbf{Gemini-2.5-Pro} repeatedly gave up rather than rephrasing or reading endpoint descriptions closely enough to recognise the equivalence (Appendix~\ref{app:sample-sessions}, Figure~\ref{fig:gemini-2_5-sprint-kickoff}). \textbf{GPT-4.1} executed several \texttt{discover} calls before finding \texttt{search\_work\_items}. It executed one search using an incorrect search string, then wrongly concluded no relevant issues existed when several did (Appendix~\ref{app:sample-sessions}, Figure~\ref{fig:gpt-4_1-security-incident-search}). 
It took \textbf{Gemini-3.1-Flash-Lite} fifteen \texttt{discover} calls to find the correct operation, causing it to ultimately exceed the 50-operation session limit (Appendix~\ref{app:sample-sessions}, Figure~\ref{fig:gemini-3_1-flash-bug-triage-round-2}). 

Importantly, this vocabulary mismatch challenge is well-known in information-retrieval \cite{furnasVocabulary1987} and not specific to Plane: any application is likely to have multiple terms in circulation for the same concept.

\paragraph{Hallucination and extra writes.} \textbf{GPT-4.1} occasionally hallucinated incorrect parameters and operation names, by guessing operations before first calling the \texttt{discover} tool to ensure accuracy (Appendix~\ref{app:sample-sessions}, Figures~\ref{fig:gpt-4_1-assign-backlog-by-specialty},~\ref{fig:gpt-4_1-incident-resp-3},~\ref{fig:gpt-4_1-in-progress-ownership-audit}). In these sessions, the hallucinations were benign: the invalid operations failed and left application state unchanged, but the same failure mode could be damaging if the agent invoked a valid operation with an incorrect name or parameters, leading to unintended side effects. \textbf{Gemini-3.1-Flash-Lite} hallucinated extra writes, producing label and assignee updates outside the scope of the prompted task (e.g., 47 extra rows in Campaign Push, 44 in Q3 Feature Prep). This is especially easy to miss outside a CoW session: an agent can satisfy the prompt while modifying unrelated application state in incorrect or harmful ways.

\subsection{Improving the tool surface}
\label{sec:improvement-iteration}

\input{figures/results-v2/run3_comparison_summary}

We updated the tool surface to address the findings, in particular to ensure ``issues'' was also used to describe operations related to ``work items''. We ran one additional trial per model; score deltas are reported in Table~\ref{tab:run3-comparison}. Models most affected by the vocabulary mismatch saw the largest gains: Gemini-2.5-Pro improved by 54\%, and GPT-4.1 by 47\%. While Gemini-3.1-Flash-Lite improved by just 3\%, it had no failed sessions in run~3 compared to at least one per iteration previously.

This loop --- CoW Scoring, localising the points of failure, and iterating or flagging known weak areas to users --- can be applied to analyze any tool surface using CoW Scoring.

\section{Limitations}

\paragraph{GT sessions are created manually.} This step takes time, but we believe the effort is comparable to building any application-specific dataset --- some investment is needed to address the construct-validity gap that motivates this work.
  
\paragraph{Evaluation assumes prompts that fully specify the intended outcome.} Real users often send underspecified prompts where multiple final states are reasonable, which CoW Scoring would conflate with failure --- custom comparators, vocabulary variation, and analyzing structural and content scores separately could partially mitigate this.
  
\paragraph{Scoring covers only write operations.} Agents also issue reads, and the volume and pattern of those reads carries diagnostic signal --- failing models in our runs often issued many redundant \texttt{discover} calls without affecting structural or content scores, but these calls in fact surfaced the work-item/issue mismatch. Incorporating read-operation signals could help diagnose these failure modes directly.

\section{Future Work}
\label{sec:future-work}

\paragraph{Input to optimization loops.} The low computational cost of CoW Scoring suggests it could also serve as a signal in optimization loops --- for example, to iteratively improve tool interfaces via GEPA \cite{agrawalGEPA2026}, or as a dense reward for model fine-tuning via GRPO-style approaches \cite{shaoDeepSeekMath2024}. Operation-level scores are particularly well-suited to this: they assign credit to partial progress towards the final world state, so optimization need not depend on the discovery of complete, successful trajectories.

\paragraph{Automated GT generation and coverage.} GT sessions are manually generated, which is inherently limited in scale and may not give uniform coverage of the API surface. Automated discovery methods --- e.g.\ MCTS-style \cite{kocsisBandit2006, zhouLanguage2024} exploration with LLM-guided candidate selection --- could generate candidate sessions, with human review reserved for auditing.

\section*{Impact Statement}

CoW Scoring aims to support safer agent deployments by making evaluations possible directly in application environments. The framework risks giving false confidence if GT sessions are unrepresentative of real usage: strong performance on a narrow set of evaluated workflows does not guarantee correct behaviour on others. Though not the focus of this paper, the CoW mechanism can also be used as a general runtime safeguard against unintended agent writes. 

\bibliographystyle{icml2026}
\bibliography{references}

\newpage
\appendix
\onecolumn

\section{Copy-on-Write (CoW) Mechanism}
\label{app:cow}

This appendix expands on the CoW mechanism summarised in Section~\ref{sec:cow}, detailing the four database-level components that together intercept and isolate agent writes without modifying production data:

\begin{enumerate}
  \item \textbf{Base tables}: These tables contain the underlying (base) production data, and follow the schema definitions specified in the application.
  \item \textbf{Changes tables}: Each base table has a corresponding changes table. Changes tables are structurally identical to base tables (with four additional columns: \texttt{session\_id}, \texttt{operation\_id}, \texttt{\_cow\_updated\_at}, \texttt{\_cow\_deleted}) but only contain rows that were written to during the agent session.
  \item \textbf{Views \cite{GarciaMolina2001}}: A SQL view is a query stored under a name. A view can be queried in the same way as a table, although no physical table exists --- querying a view instead executes the underlying query (the view definition) and returns its results, so it can be read from like a table. In the CoW mechanism, views are used to merge each base table with its corresponding changes table, so that reads during an agent session return rows from the base table, with any rows that the agent has modified in the current session replaced by their changes-table versions.
  \item \textbf{Triggers \cite{GarciaMolina2001}}: \texttt{INSTEAD OF} triggers can be defined on views, and define how a given operation (in this case, a database write) should be carried out. For CoW, writes should be forwarded to the changes table. 
  \begin{enumerate}[label=\roman*.]\hbadness=10000\relax
    \item Create copies of the affected row(s);
    \item Apply the write to the row(s);
    \item Append the \texttt{session\_id}, \texttt{operation\_id}, \texttt{\_cow\_updated\_at} (the current time), and \texttt{\_cow\_deleted} (\texttt{True} if the write is a \texttt{DELETE} operation);
    \item Write the row(s) to the corresponding changes table.
  \end{enumerate}
  Notably, the handling of complex trigger configurations such as cascade triggers and PostgreSQL-specific triggers should be explored in future work to extend the real-world portability of the CoW mechanism.
\end{enumerate}

Enabling CoW on a database creates each of these components, and deploys the necessary trigger functions -- following Algorithm~\ref{alg:cow-enable} below. 

\begin{algorithm}[H]
\caption{Enabling CoW on a database}
\label{alg:cow-enable}
\begin{algorithmic}[1]
\FOR{each table $T$ in the schema}
  \STATE Rename $T \to T\texttt{\_base}$
  \STATE Create $T\texttt{\_changes}$ with $T$'s columns plus CoW metadata
  \STATE Create view $T$ merging $T\texttt{\_base}$ and $T\texttt{\_changes}$
  \STATE Attach \texttt{INSTEAD OF} triggers to view $T$
\ENDFOR
\STATE Deploy CoW trigger functions to the database
\end{algorithmic}
\end{algorithm}

\section{CoW Scoring}
\label{app:cow-scoring}

\begin{figure*}[t]
  \centering
  \vspace{0.4em}
  \includegraphics[width=0.95\textwidth]{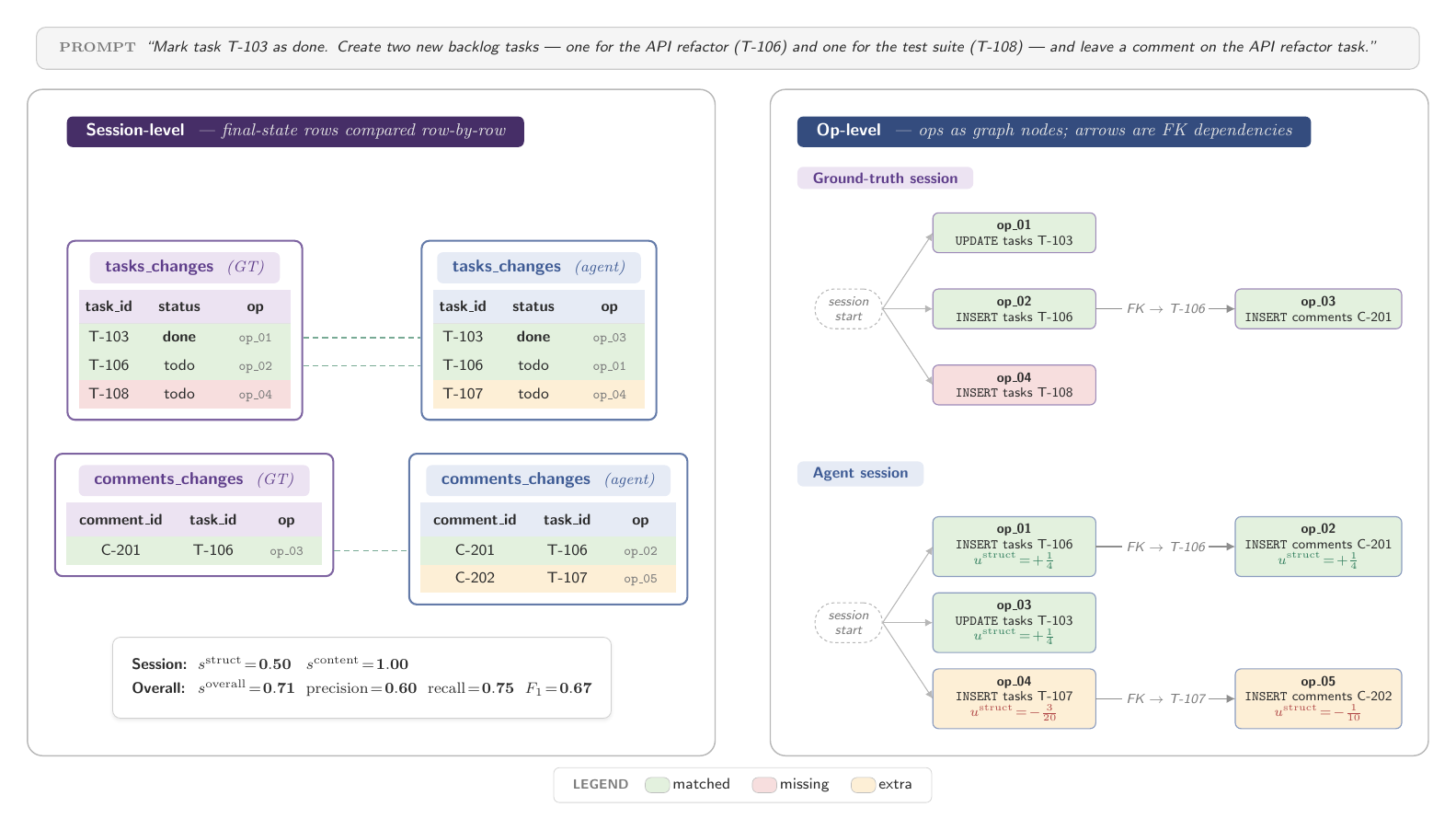}
  \vspace{0.4em}
  \caption{\textbf{Worked example of CoW scoring.} Session-level scores ($s^{\mathrm{struct}}$, $s^{\mathrm{content}}$) are calculated by comparing all rows changed for a given GT and agent session pair (\textbf{left}); and operation-level scores ($u^{\mathrm{struct}}(o_i)$, $u^{\mathrm{content}}(o_i)$) are calculated using the $\Delta$ in session-level structural score, and the content similarity of the rows written by each operation in the agent session (\textbf{right}).}
  \label{fig:scoring-example}
\end{figure*}

Figure~\ref{fig:scoring-example} provides a worked example of the scoring procedure described in Section~\ref{sec:scoring}, illustrating how the session- and operation-level scores are computed from a paired GT and agent session. On the left, rows are colour-coded as matched, missing, and extra -- which is used to compute $s^{\mathrm{struct}}$ and $s^{\mathrm{content}}$. On the right, each agent operation $o_i$ is attributed a structural utility $u^{\mathrm{struct}}(o_i)$ from the change in session-level structural score, and a content utility $u^{\mathrm{content}}(o_i)$ from the similarity of the rows it wrote.

\subsection{\texttt{agent-cow} Scoring Library}
\label{app:cow-scoring-lib}

\begin{figure}[t]
  \centering
  \vspace{0.6em}
  \includegraphics[width=\columnwidth]{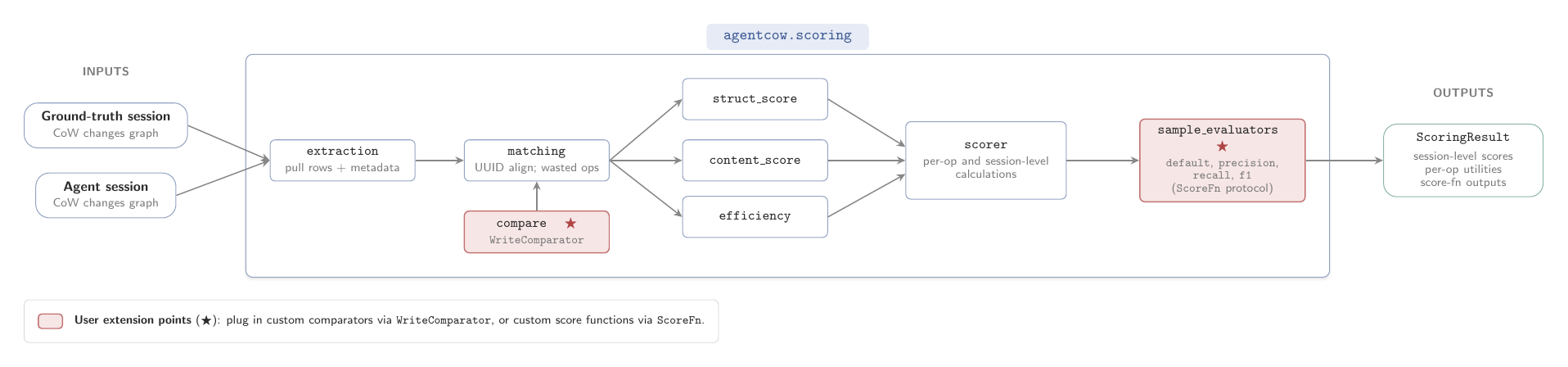}
  \vspace{0.8em}
  \caption{
    \textbf{Copy-on-Write (CoW) scoring library architecture.}
  }
  \label{fig:cow-scoring-lib}
\end{figure}

The \texttt{agent-cow} scoring module (visualized in Figure~\ref{fig:cow-scoring-lib}) scores an agent's CoW session against a ground-truth recording via the API entry point \texttt{score\_cow\_sessions}. The pipeline has five stages:

\begin{enumerate}
  \item \textbf{Extraction} (\texttt{extraction.py}) --- rows are read from \texttt{*\_changes} tables, grouped by \texttt{operation\_id}, and sorted by \texttt{\_cow\_updated\_at}.
  \item \textbf{Matching} (\texttt{matching.py}) --- both sides are reduced to one row per \texttt{(table, pk)} (last write wins). Each ground-truth entity greedily picks the best matching agent entity in the same table; a UUID mapping bridges the different primary keys created by each side.
  \item \textbf{Field comparison} (\texttt{compare.py}) --- each field is compared by SQL type: text uses \texttt{SequenceMatcher.ratio()}, JSON is deep-equal, everything else is exact. PKs, FKs (remapped via the UUID mapping), timestamps, and configured \texttt{ignored\_fields} are excluded from content scoring.
  \item \textbf{Per-op scoring} (\texttt{scorer.py}) --- for each agent operation in topological order the cumulative agent rows are re-scored against ground truth, recording the delta in \texttt{op\_struct\_scores}. \texttt{op\_content\_scores} is computed independently from the final matching.
  \item \textbf{Reduce} (\texttt{scores.py}) --- registered \texttt{score\_fns} reduce the \texttt{ScoringResult} to scalar entries on \texttt{result.scores}.
\end{enumerate}

\subsubsection{Scoring Customization}
\label{app:customization}

\paragraph{Comparators.} The default content similarity uses one comparator per SQL column data type. To override comparison for a specific table, callers can supply either a row-level function --- which receives the GT and agent rows as plain dicts (with FK UUIDs already remapped into GT space), returns a bool or float in $[0, 1]$ --- or a \texttt{WriteComparator}, which additionally exposes the row's CoW metadata, the table's column-type metadata, and the cross-session UUID mapping for full control. Additionally, passing \texttt{collapse=True} drops entities the agent created and later deleted within the session before scoring, reflecting only the agent's net intent.

\paragraph{Score functions.} Any callable mapping the raw scoring terms to a float can be registered as an \texttt{overall} score. The \texttt{agent-cow} library includes sample scoring functions, each mapping a \texttt{ScoringResult} to a single float, which can be passed via the \texttt{score\_fns} kwarg, and their outputs are written to \texttt{result.scores} under the names below. 

\paragraph{Default overall score.} An average of the two session-level signals:
\begin{equation}
  s^{\mathrm{overall}} = 0.5 \cdot s^{\mathrm{struct}} + 0.5 \cdot s^{\mathrm{content}}
\end{equation}

\paragraph{Precision.} Of the rows the agent wrote, the fraction that matched a GT entity:
\begin{equation}
  \mathrm{precision} = \frac{N_{\mathrm{matched}}}{N_{\mathrm{matched}} + N_{\mathrm{extra}}}
\end{equation}

\paragraph{Recall.} Of the rows GT session wrote, the fraction the agent reproduced:
\begin{equation}
  \mathrm{recall} = \frac{N_{\mathrm{matched}}}{N_{\mathrm{matched}} + N_{\mathrm{missing}}}
\end{equation}

\paragraph{F1.} Harmonic mean of precision and recall:
\begin{equation}
  F_1 = \frac{2 \cdot \mathrm{precision} \cdot \mathrm{recall}}{\mathrm{precision} + \mathrm{recall}}
\end{equation}

\section{Plane Integration and Experimental Setup}
\label{app:plane-integration}

\subsection{CoW Integration in Plane}
\label{app:cow-plane}

The full implementation of CoW into Plane can be found in the following GitHub repository: \href{https://github.com/JoannaRoy/plane-cow}{Plane (with CoW)}, which also includes the raw scoring results (\href{https://github.com/JoannaRoy/plane-cow/blob/preview/results.zip}{\texttt{results.zip}}). 

The breakdown of application-side additions is shown in Table~\ref{tab:cow-integration}. The far-right column indicates whether that portion of code could be ported to the harness if the user wanted to simplify their implementation. Notably, the recording API and CoW scoring metadata (session IDs, scores, API versions, prompts) were included in the Plane application for this experiment, but could be stored in the harness for a simpler, less-invasive implementation. Only the CoW mechanism (using the \texttt{agent-cow-python} library) and associated deployment commands are necessary.

\input{figures/cow_integration_table}

\subsection{Workspace Initial Data Seeding}
\label{app:seeding}

The Plane workspace was seeded with the entities summarised in Table~\ref{tab:exp-seed-data} prior to any agent runs. Ground-truth CoW recordings were then captured against this fixed initial state.

\input{figures/datasets/seed_data_table}

\subsection{Ground-Truth Session Setup}
\label{app:gt-sessions}

To produce a realistic and diverse set of ground-truth (GT) workflows without spending excessive author effort, an LLM agent (Claude Opus 4.6) under human supervision was used. The agent was given read access to the \texttt{plane-cow} codebase --- including the Plane data model, the seeded workspace dump (Appendix~\ref{app:seeding}), and the CoW recording API --- so it could ground its workflows in the actual entities present in the codebase and database. 

We provided the agent with short workflow themes (e.g.\ \emph{sprint kickoff}, \emph{security incident response}, \emph{stale backlog cleanup}) drawn from the kinds of bulk operations someone might be expected to perform in a project management tool. The agent then started a CoW recording, and executed the corresponding sequence of API writes against the seeded workspace, and drafted a natural-language prompt expressing the workflow.

Each candidate session was then reviewed. We checked that (i) the prompt was unambiguous and self-contained, (ii) every recorded operation was justified by the prompt, (iii) no operation was missing relative to a literal reading of the prompt.

\begin{figure}[H]
  \centering
  \includegraphics[width=0.78\textwidth]{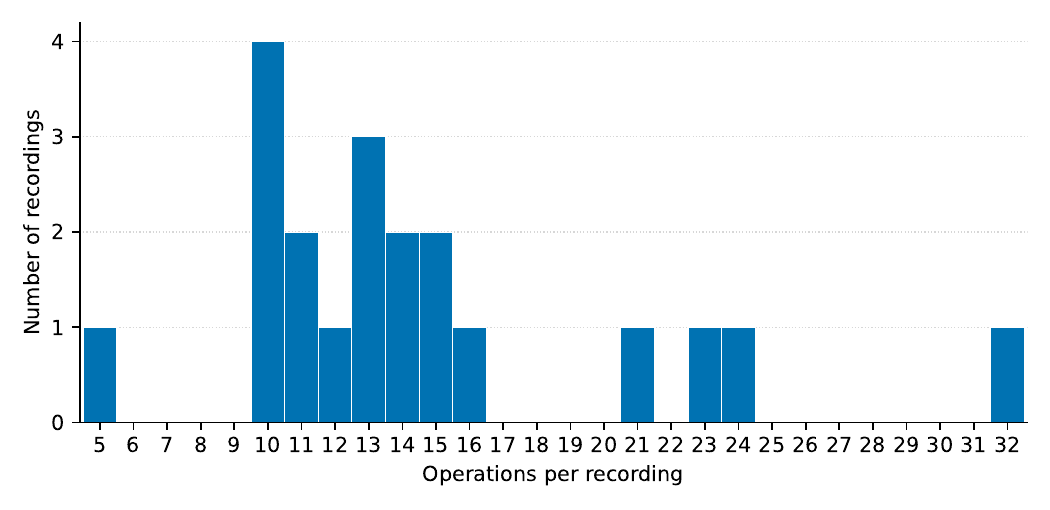}
  \caption{Distribution of write operations per ground-truth (GT) session across the 20 workflows.}
  \label{fig:gt-op-distribution}
\end{figure}

In a real application, the product team would likely have some key workflows they would be interested in having the agent execute and perform well on, which would then be recorded manually.

\input{figures/datasets/recordings_table.tex}

\section{Results Summary}
\label{app:results-summary}

This appendix collects per-model summaries from the preliminary study (Section~\ref{sec:results}). The raw scoring data for all 300 sessions is available at \href{https://github.com/JoannaRoy/plane-cow/blob/preview/results.zip}{\texttt{results.zip}}. Table~\ref{tab:results-summary} reports the overall score $s^{\mathrm{overall}}$ (Appendix~\ref{app:customization}) for each model averaged across the first two trials, alongside per-trial scores and row-level precision, recall, and F1 derived from the matched/missing/extra counts. Table~\ref{tab:worst-sessions} lists the five lowest-scoring (model, workflow) pairs per model; these are the sessions inspected in Section~\ref{sec:results} to surface the failure modes discussed in the main text and illustrated in Appendix~\ref{app:sample-sessions}.

\input{figures/results-v1/results_summary_table}

\input{figures/results-v1/worst_sessions_combined_table}

\section{Sample Sessions}
\label{app:sample-sessions}

\begin{figure}[H]
\begin{minipage}[t]{0.48\textwidth}
  \centering
  \includegraphics[width=\textwidth]{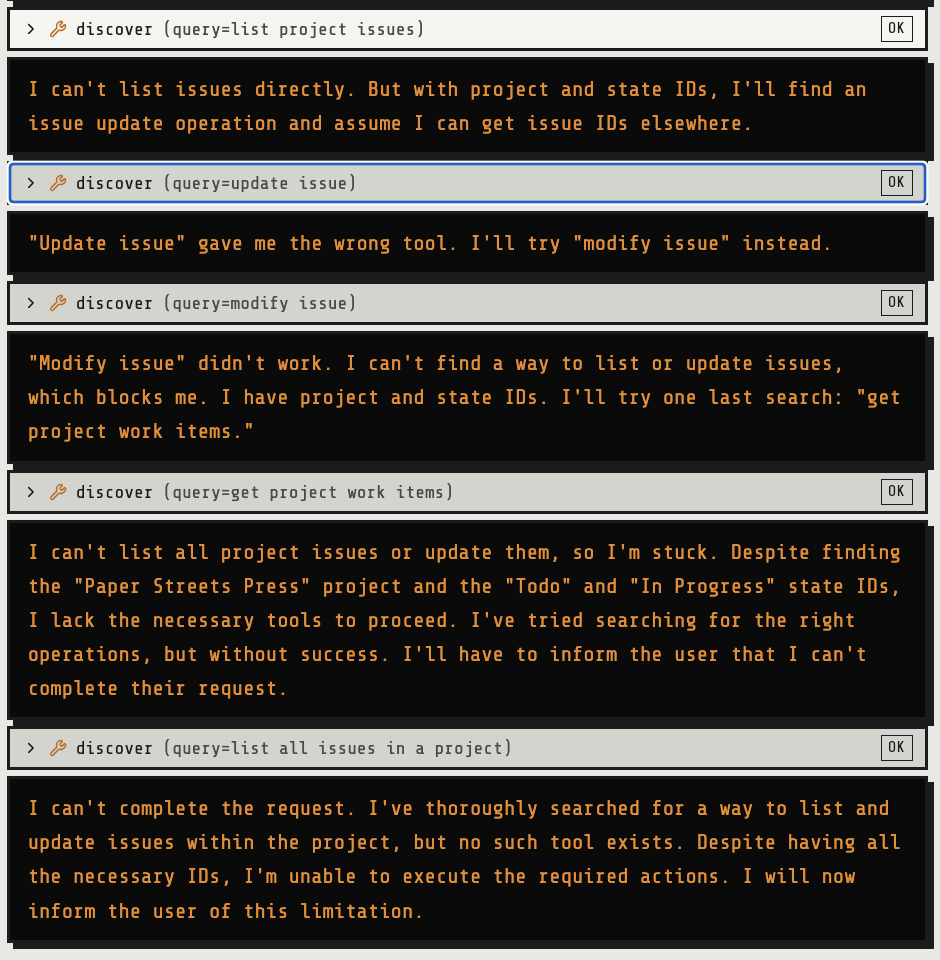}
  \caption{Gemini-2.5-Pro tool calls for the Sprint Kickoff task. The model gives up after its first \texttt{discover} call returns no matching endpoints for ``issues''.}
  \label{fig:gemini-2_5-sprint-kickoff}
\end{minipage}\hfill
\begin{minipage}[t]{0.48\textwidth}
  \centering
  \includegraphics[width=\textwidth]{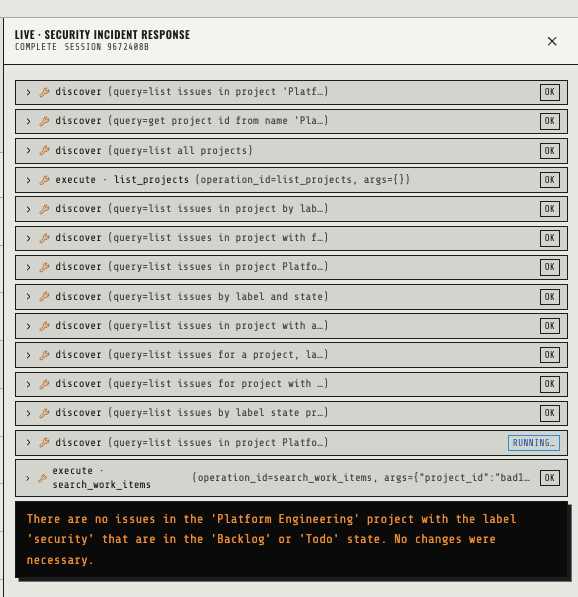}
  \caption{GPT-4.1 tool calls for the Security Incident Response task. The model issues several \texttt{discover} calls before finding \texttt{search\_work\_items}, then queries with an unsupported search string format, receives an empty result, and concludes no relevant issues exist.}
  \label{fig:gpt-4_1-security-incident-search}
\end{minipage}
\end{figure}

\begin{figure}[H]
\begin{minipage}[t]{0.48\textwidth}
  \centering
  \includegraphics[width=\textwidth]{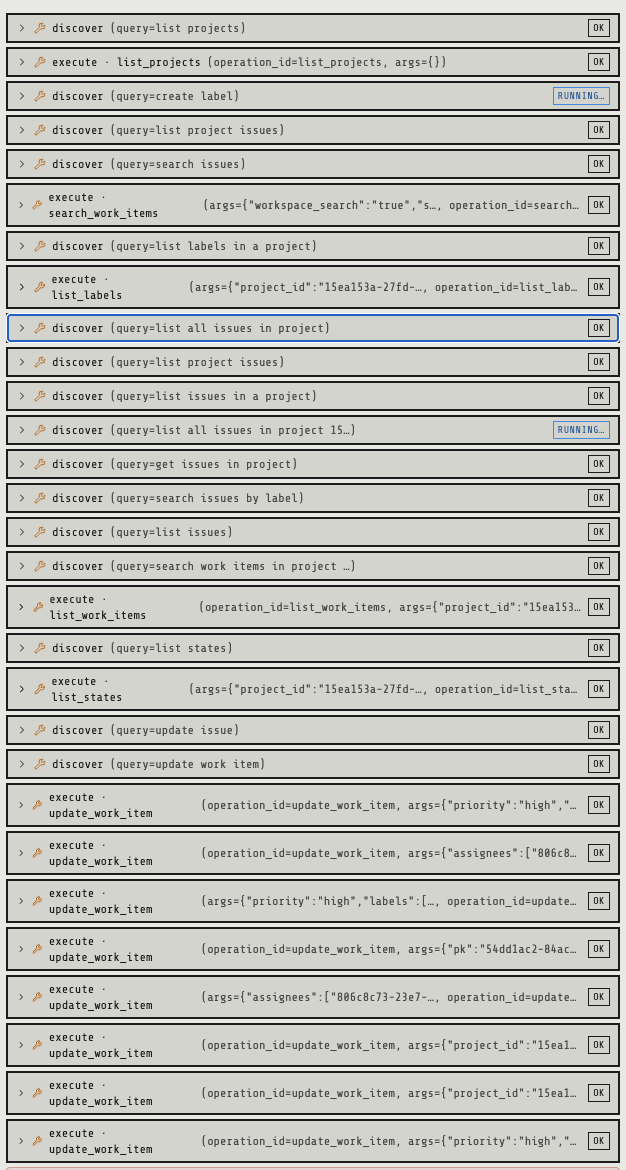}
  \caption{Gemini-3.1-Flash-Lite tool calls for the Bug Triage Round 2 task. The model issues over fifteen \texttt{discover} calls before finding the correct endpoint, exhausting the 50-operation limit.}
  \label{fig:gemini-3_1-flash-bug-triage-round-2}
\end{minipage}\hfill
\begin{minipage}[t]{0.48\textwidth}
  \centering
  \includegraphics[width=\textwidth]{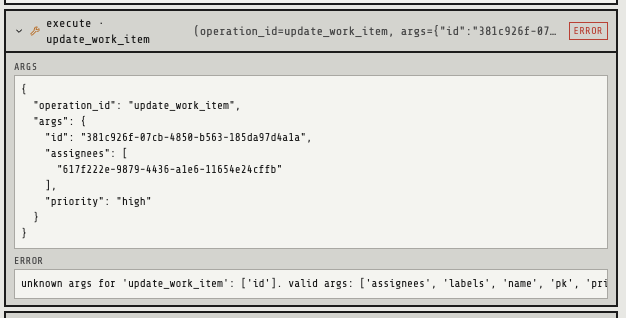}
  \caption{GPT-4.1 tool calls for the Assign Backlog by Specialty task. The model executes \texttt{update\_work\_item} without first calling \texttt{discover}, passing \texttt{id} as an argument; the API rejects it since the valid parameter is \texttt{pk}.}
  \label{fig:gpt-4_1-assign-backlog-by-specialty}
\end{minipage}
\end{figure}

\begin{figure}[H]
\begin{minipage}[t]{0.48\textwidth}
  \centering
  \includegraphics[width=\textwidth]{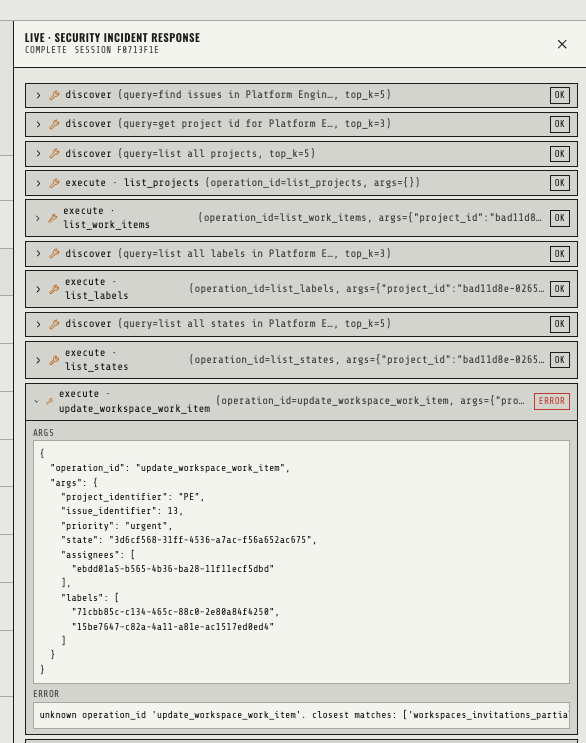}
  \caption{GPT-4.1 tool calls for the Security Incident Response task. The model executes \texttt{update\_workspace\_work\_item} without first calling \texttt{discover}, guessing an \texttt{operation\_id} that does not exist in the API.}
  \label{fig:gpt-4_1-incident-resp-3}
\end{minipage}\hfill
\begin{minipage}[t]{0.48\textwidth}
  \centering
  \includegraphics[width=\textwidth]{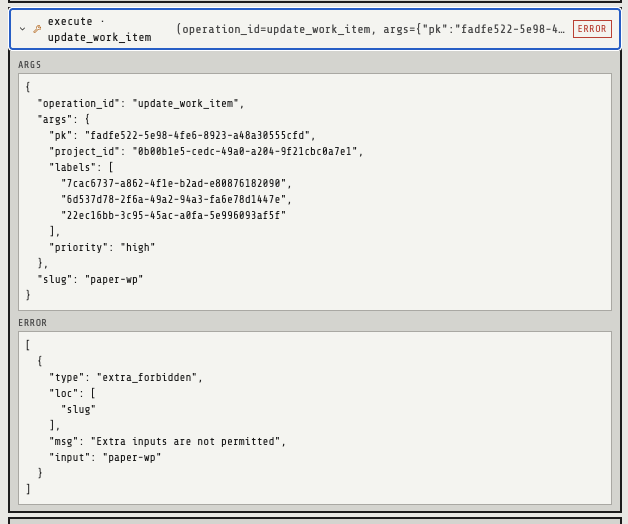}
  \caption{GPT-4.1 tool calls for the In-Progress Ownership Audit task. The model calls \texttt{update\_work\_item} with an extra \texttt{slug} field not accepted by the API, resulting in an \texttt{extra\_forbidden} validation error.}
  \label{fig:gpt-4_1-in-progress-ownership-audit}
\end{minipage}
\end{figure}

This appendix contains tool-call traces from representative low-scoring sessions identified in Table~\ref{tab:worst-sessions}, which illustrate the failure modes discussed in Section~\ref{sec:results}: vocabulary mismatch between prompt terminology and the API surface (Figures~\ref{fig:gemini-2_5-sprint-kickoff},~\ref{fig:gpt-4_1-security-incident-search}, and~\ref{fig:gemini-3_1-flash-bug-triage-round-2}), and operations issued without a prior \texttt{discover} call to obtain the correct argument format (Figures~\ref{fig:gpt-4_1-assign-backlog-by-specialty}, ~\ref{fig:gpt-4_1-incident-resp-3} and~\ref{fig:gpt-4_1-in-progress-ownership-audit}).  Despite the overall improvement from runs~1 and 2 to run~3, two model-specific failure modes persisted across iterations:

\begin{itemize}
  \item \textbf{GPT-4.1:} Occasionally executes an operation without first calling \texttt{discover} to obtain the correct argument format, producing API errors (Appendix~\ref{app:sample-sessions}, Figures~\ref{fig:gpt-4_1-assign-backlog-by-specialty} and~\ref{fig:gpt-4_1-in-progress-ownership-audit}) and one failed session in run~3.
  \item \textbf{Gemini-3.1-Flash-Lite:} Continued to write extra rows. For example, in the sessions: Sprint Kickoff (8 extra work items and assignees), Deprioritize Stale Medium Backlog (5 extra labels), and Infra Sprint Kickoff (18 extra labels and assignees), none of which were prompted.
\end{itemize}

\end{document}

%% file: figures/results-v2/run3_comparison_summary.tex
\begin{table}[t]
\centering
\caption{
  \textbf{Run 3 vs.\ runs 1 \& 2 mean overall score.}
  Run~3 used an updated tool surface based off the analysis of runs~1 and 2.
}
\label{tab:run3-comparison}
\resizebox{\columnwidth}{!}{%
\begin{tabular}{lcccc}
\toprule
Model & Avg(1,2) & Run\,3 & $n$ & $\Delta$ \\
\midrule
\texttt{GPT-4.1} & 0.32 & 0.79 & 20 & \textbf{+0.47} \\
\texttt{Gemini 2.5 Pro} & 0.43 & 0.98 & 20 & \textbf{+0.54} \\
\texttt{Gemini 3.1 Flash-Lite} & 0.80 & 0.83 & 20 & +0.03 \\
\texttt{GPT-5} & 0.95 & 0.97 & 20 & +0.02 \\
\texttt{Gemini 3.1 Pro} & 0.96 & 0.98 & 20 & +0.01 \\
\bottomrule
\end{tabular}%
}
\end{table}

%% file: figures/cow_integration_table.tex
\begin{table}[H]
  \centering
  \small
  \begin{tabularx}{\columnwidth}{@{}p{3.2cm}r>{\raggedright\arraybackslash}Xl@{}}
  \toprule
  \textbf{Category} & \textbf{Lines} & \textbf{Contents} & \textbf{Portable?} \\
  \midrule
  (a) Core CoW machinery
    & \textasciitilde250
    & Middleware that activates a CoW session by reading the \texttt{session\_id}
      and \texttt{operation\_id} headers on every agent request.
      \par\smallskip
      Endpoints to commit or discard a session's writes, check whether CoW is
      enabled, and return the operation IDs of a session (the endpoint the
      harness queries to retrieve session state for scoring).
      \par\smallskip
      Plane-specific list of tables to exclude from the CoW mechanism.
    & No \\
  \addlinespace
    (b) Deployment commands
      & \textasciitilde180
      & One-time installation of the SQL functions, views, and triggers described
        in Section~\ref{sec:cow-scoring} into the application database.
        \par\smallskip
        Commands that enable, disable, and re-enable CoW.
      & No \\
  \addlinespace
  (c) Recording API --- GT \& agent session tracking
    & \textasciitilde680
    & CRUD endpoints for GT recordings and agent runs (their prompts, model
      metadata, and links between the two).
      \par\smallskip
      Plane-specific scoring configuration --- which tables and fields to ignore when
      comparing sessions --- together with helpers for writing scoring results to
      CSV / JSONL.
    & Yes \\
  \midrule
  \textbf{Total} & \textbf{\textasciitilde790} & \multicolumn{1}{l}{} & \\
  \bottomrule
  \end{tabularx}
  \vspace{4pt}
  \caption{Application-side additions required to integrate \texttt{agent-cow}
  into Plane. Category~(a) is the minimum any Django project would need to
  replicate; categories~(b) and~(c) are specific to our evaluation setup and
  could move into the eval harness.}
  \label{tab:cow-integration}
\end{table}

%% file: figures/datasets/seed_data_table.tex
\begin{table}[H]
  \centering
  \small
  \begin{tabularx}{\columnwidth}{@{}p{2.2cm}>{\raggedright\arraybackslash}Xrrrr@{}}
  \toprule
  \textbf{Entity} & \textbf{Description} & \textbf{PSP} & \textbf{PE} & \textbf{GS} & \textbf{Total} \\
  \midrule
  Projects
    & Top-level Plane projects covering distinct organisational domains.
    & --- & --- & --- & 3 \\
  \addlinespace
  Members
    & Workspace users (one admin agent plus six personas with role-aligned profiles).
    & --- & --- & --- & 7 \\
  \addlinespace
  States
    & Per-project workflow states (\texttt{Backlog}, \texttt{Todo}, \texttt{In Progress}, \texttt{Done}, \texttt{Cancelled}).
    & 5 & 5 & 5 & 15 \\
  \addlinespace
  Labels
    & Per-project tag taxonomy (e.g.\ \texttt{bug}, \texttt{security}, \texttt{urgent}).
    & 10 & 8 & 8 & 26 \\
  \addlinespace
  Work items
    & Seeded issues with assignees, priorities, labels, and descriptions.
    & 32 & 16 & 15 & 63 \\
  \bottomrule
  \end{tabularx}
  \vspace{4pt}
  \caption{Entities seeded into the Plane workspace before any agent runs. Project-level resources are reported per project (PSP = Paper Streets Press, PE = Platform Engineering, GS = Growth and Subscriptions);
  workspace-level resources only have a workspace total.}
  \label{tab:exp-seed-data}
\end{table}

%% file: figures/datasets/recordings_table.tex
{\small
\setlength{\LTpre}{6pt}\setlength{\LTpost}{6pt}
\begin{longtable}{@{}r >{\raggedright\arraybackslash}p{3.4cm} l r >{\raggedright\arraybackslash}p{9.4cm}@{}}
  \caption{Ground-truth CoW recordings used in evaluation. \emph{Ops} is the number of mutating API calls captured in the recording. Projects: PSP = Paper Streets Press, PE = Platform Engineering, GS = Growth and Subscriptions.}\label{tab:gt-recordings} \\
  \toprule
  \textbf{\#} & \textbf{Recording name} & \textbf{Project} & \textbf{Ops} & \textbf{Prompt} \\
  \midrule
  \endfirsthead
  \multicolumn{5}{c}{\tablename\ \thetable{} -- continued from previous page} \\
  \toprule
  \textbf{\#} & \textbf{Recording name} & \textbf{Project} & \textbf{Ops} & \textbf{Prompt} \\
  \midrule
  \endhead
  \midrule
  \multicolumn{5}{r}{\textit{continued on next page}} \\
  \endfoot
  \midrule
  \textbf{Total} & & & \textbf{292} & \\
  \bottomrule
  \endlastfoot
  1 & Sprint Kickoff & PSP & 12 & In project 'Paper Streets Press', move every issue in state 'Todo' to state 'In Progress'. For each that is currently unassigned, assign @alice (user id: 2d02e7e3-e5be-4687-8fbf-32a133b3b271). For each that is currently assigned to @blub (user id: fad82a69-23ed-4260-83f6-cae2caf466a5), reassign to @bob (user id: 617f222e-9879-4436-a1e6-11654e24cffb). \\
  \addlinespace[2pt]
  2 & Security Incident Response & PE & 13 & In project 'Platform Engineering', find every issue with label 'security' that is in state 'Backlog' or 'Todo'. Move each to state 'In Progress'. Add label 'blocker' if it does not already have it. Set priority to 'urgent' if not already urgent. Assign @dave (user id: ebdd01a5-b565-4b36-ba28-11f11ecf5dbd) if currently unassigned. \\
  \addlinespace[2pt]
  3 & Bug Escalation & GS & 15 & In project 'Growth and Subscriptions', find every issue with label 'bug' currently in state 'Backlog'. Move each to state 'Todo'. Add label 'blocker' to each. Assign @frank (user id: 806c8c73-23e7-42f9-9ed4-b9e3cb8c7bfd) to each. \\
  \addlinespace[2pt]
  4 & Stale Backlog Cleanup & PSP & 10 & In project 'Paper Streets Press', find every issue in state 'Backlog' that is unassigned and has priority 'low' or 'none'. Add label 'stale' to each and transition it to state 'Cancelled'. \\
  \addlinespace[2pt]
  5 & Triage Unassigned Backlog & GS & 11 & In project 'Growth and Subscriptions', find every issue in state 'Backlog' with no assignee. Assign @eve (45a64208-112f-4bd4-a284-481cab326106) to issues with priority 'high' or 'urgent'. Assign @frank (806c8c73-23e7-42f9-9ed4-b9e3cb8c7bfd) to issues with priority 'medium'. For issues with priority 'low' or 'none', set priority to 'medium' and assign @frank. \\
  \addlinespace[2pt]
  6 & Stale Backlog Cancel & PE & 10 & In project 'Platform Engineering', find every issue in state 'Backlog' with no assignee, priority 'low' or 'medium', and without label 'security'. Add label 'stale' to each and transition it to state 'Cancelled'. \\
  \addlinespace[2pt]
  7 & Assign Backlog by Specialty & PE & 14 & In project 'Platform Engineering', find every issue in state 'Backlog' with no assignee. Assign @dave (ebdd01a5) to issues with label 'security'. Assign @bob (617f222e) to issues with label 'bug'. Assign @carol (71bb681c) to all other unassigned Backlog issues. For each newly assigned issue with priority 'medium' or lower, also set priority to 'high'. \\
  \addlinespace[2pt]
  8 & Deprioritize Stale Medium Backlog & PSP & 10 & In project 'Paper Streets Press', find every issue in state 'Backlog' with priority 'medium' and no assignee. Set each to priority 'low' and add label 'stale'. \\
  \addlinespace[2pt]
  9 & Escalate High to Urgent & PE & 11 & In project 'Platform Engineering', find every issue with priority 'high' in any open state (Backlog, Todo, In Progress). Set each to priority 'urgent'. For those in state 'Backlog', also add label 'urgent'. \\
  \addlinespace[2pt]
  10 & Assign Backlog by Role & PSP & 15 & In project 'Paper Streets Press', find every issue in state 'Backlog' with no assignee. Assign @alice (2d02e7e3) to issues with label 'bug'. Assign @carol (71bb681c) to issues with label 'security'. Assign @bob (617f222e) to all other unassigned Backlog issues. \\
  \addlinespace[2pt]
  11 & Q3 Feature Prep & GS & 24 & In project 'Growth and Subscriptions', create a label 'q3-target-v2' with color \#8b5cf6. Find every issue with label 'growth' or 'experiment' that is not in state Done or Cancelled. Add the 'q3-target-v2' label to each (preserving existing labels). Move each from Backlog to Todo if currently in Backlog. Assign @eve (45a64208-112f-4bd4-a284-481cab326106) to any that are unassigned. Set priority to 'high' for any with priority none, low, or medium. \\
  \addlinespace[2pt]
  12 & Infra Sprint Kickoff & PE & 23 & In project 'Platform Engineering', create a label 'infra-sprint-v2' with color \#0284c7. Find every issue with label 'infra' that is not in state Done or Cancelled. Add the 'infra-sprint-v2' label to each (preserving existing labels). Move each from Backlog to Todo if currently in Backlog. Assign @dave (ebdd01a5-b565-4b36-ba28-11f11ecf5dbd) to any that are unassigned. Set priority to 'high' for any with priority none, low, or medium. \\
  \addlinespace[2pt]
  13 & Bug Triage Round 2 & GS & 16 & In project 'Growth and Subscriptions', create a label 'bug-sprint-v2' with color \#dc2626. Find every issue with label 'bug' that is not in state Done or Cancelled. Add the 'bug-sprint-v2' label to each (preserving existing labels). Also add the 'blocker' label if not already present. Set priority to 'high' for any with priority none, low, or medium. Assign @frank (806c8c73-23e7-42f9-9ed4-b9e3cb8c7bfd) to any that are unassigned. \\
  \addlinespace[2pt]
  14 & Campaign Push & GS & 5 & In project 'Growth and Subscriptions', create a label 'campaign-active-v2' with color \#db2777. Find every issue with label 'campaign' that is not in state Done or Cancelled. Add the 'campaign-active-v2' label to each (preserving existing labels). Move each from Backlog to Todo if currently in Backlog. For issues with priority 'high', set priority to 'urgent'. Assign @eve (45a64208-112f-4bd4-a284-481cab326106) to any that are unassigned. \\
  \addlinespace[2pt]
  15 & In Progress Ownership Audit & PSP & 10 & In project 'Paper Streets Press', create a label 'active-sprint-v2' with color \#2563eb. Find every issue currently in state 'In Progress'. Add the 'active-sprint-v2' label to each (preserving existing labels, but removing the 'stale' label if present since in-progress work is no longer stale). Assign @alice (2d02e7e3-e5be-4687-8fbf-32a133b3b271) to any that are unassigned. Set priority to 'high' for any with priority none, low, or medium. \\
  \addlinespace[2pt]
  16 & Editorial Sprint & PSP & 13 & In project 'Paper Streets Press', create a label 'editorial-sprint-v2' with color \#f59e0b. Find every issue with label 'editorial' that is not in state Done or Cancelled. Add the 'editorial-sprint-v2' label to each (preserving existing labels). Move each from Backlog to Todo if currently in Backlog. Assign @alice (2d02e7e3-e5be-4687-8fbf-32a133b3b271) to any that are unassigned. Set priority to 'high' for any with priority none, low, or medium. \\
  \addlinespace[2pt]
  17 & Performance Backlog Push & PE & 13 & In project 'Platform Engineering', create a label 'perf-sprint-v2' with color \#7c3aed. Find every issue with label 'performance' that is in state Backlog or Todo. Add the 'perf-sprint-v2' label to each (preserving existing labels). Set priority to 'high' for any with priority none, low, or medium. Assign @carol (71bb681c-bace-4ff8-834c-9b169cd6bfc9) to any that are unassigned. Move each from Backlog to Todo if currently in Backlog. \\
  \addlinespace[2pt]
  18 & Bug Fix Sprint & PSP & 21 & In project 'Paper Streets Press', create a label 'bug-fix-sprint-v2' with color \#b91c1c. Find every issue with label 'bug' that is not in state Done or Cancelled. Add the 'bug-fix-sprint-v2' label to each (preserving existing labels). Set priority to 'high' for any with priority none, low, or medium. Move each from Backlog to Todo if currently in Backlog. Assign @bob (617f222e-9879-4436-a1e6-11654e24cffb) to any that are unassigned. \\
  \addlinespace[2pt]
  19 & Security Hardening Sprint & PE & 14 & In project 'Platform Engineering', create a label 'security-sprint-v2' with color \#991b1b. Find every issue with label 'security' that is not in state Done or Cancelled. Add the 'security-sprint-v2' label to each (preserving existing labels). Also ensure the 'blocker' label is present on each. Set priority to 'urgent' for any not already urgent. Move each from Backlog to Todo if currently in Backlog. Assign @dave (ebdd01a5-b565-4b36-ba28-11f11ecf5dbd) to any that are unassigned. \\
  \addlinespace[2pt]
  20 & Backlog Ownership Assignment & GS & 32 & In project 'Growth and Subscriptions', create a label 'needs-owner' with color \#d97706. Find every issue in state Backlog with no assignee. Add the 'needs-owner' label to each (preserving existing labels). Assign @eve (45a64208-112f-4bd4-a284-481cab326106) to those with priority 'high' or 'urgent'. Assign @frank (806c8c73-23e7-42f9-9ed4-b9e3cb8c7bfd) to those with priority 'medium'. For issues with priority 'none' or 'low', set priority to 'medium' then assign @frank. Move each issue to Todo state. \\
\end{longtable}
}

%% file: figures/results-v1/results_summary_table.tex
\begin{table}[t]
\centering
\caption{
  \textbf{Per-model summary across trials per model.}
  Overall ($s^{\mathrm{overall}}$) score as defined in Appendix~\ref{app:customization}.
  Precision, recall, and F1 are computed from row-level matched/missing/extra counts.
}
\label{tab:results-summary}
\resizebox{\columnwidth}{!}{%
\begin{tabular}{lccccccc}
\toprule
Model & Completion & \multicolumn{3}{c}{Overall ($s^{\mathrm{overall}}$)} & Precision & Recall & F1 \\
\cmidrule(lr){3-5}
 &  & Average & Run\,1 & Run\,2 &  &  &  \\
\midrule
\texttt{gpt-4.1} & 40/40 & 0.32 & 0.31 & 0.33 & 0.92 & 0.16 & 0.28 \\
\texttt{gemini-2.5-pro} & 40/40 & 0.43 & 0.45 & 0.42 & 0.98 & 0.21 & 0.35 \\
\texttt{gemini-3.1-flash-lite} & 38/40 & 0.80 & 0.83 & 0.77 & 0.78 & 0.83 & 0.81 \\
\texttt{gpt-5} & 40/40 & 0.95 & 0.97 & 0.92 & 0.98 & 0.98 & 0.98 \\
\texttt{gemini-3.1-pro} & 40/40 & 0.96 & 0.96 & 0.97 & 0.96 & 1.00 & 0.98 \\
\bottomrule
\end{tabular}%
}
\end{table}

%% file: figures/results-v1/worst_sessions_combined_table.tex
\begin{table*}[ht]
  \centering
  \small
  \resizebox{\textwidth}{!}{%
  \begin{tabular}{r l r l r l r l r l r}
    \toprule
     & \multicolumn{2}{c}{GPT-4.1} & \multicolumn{2}{c}{Gemini 2.5 Pro} & \multicolumn{2}{c}{Gemini 3.1 Flash Lite} & \multicolumn{2}{c}{GPT-5} & \multicolumn{2}{c}{Gemini 3.1 Pro} \\
    \cmidrule(lr){2-3} \cmidrule(lr){4-5} \cmidrule(lr){6-7} \cmidrule(lr){8-9} \cmidrule(lr){10-11}
    \# & Session & \% & Session & \% & Session & \% & Session & \% & Session & \% \\
    \midrule
    1 & Bug Escalation & 0.0 & Sprint Kickoff & 0.0 & Stale Backlog Cleanup & 56.7 & Stale Backlog Cancel & 50.0 & Security Hardening Sprint & 87.5 \\
    2 & Stale Backlog Cleanup & 0.0 & Bug Escalation & 0.0 & Stale Backlog Cancel & 59.0 & Security Incident Response & 84.5 & Security Incident Response & 89.1 \\
    3 & Security Incident Response & 0.0 & Security Incident Response & 0.0 & Q3 Feature Prep & 63.0 & Deprioritize Stale Med.\ Backlog & 95.8 & Bug Triage Round 2 & 91.5 \\
    4 & Stale Backlog Cancel & 0.0 & Assign Backlog by Role & 0.0 & Bug Fix Sprint & 66.7 & Assign Backlog by Specialty & 95.8 & Bug Fix Sprint & 93.2 \\
    5 & Triage Unassigned Backlog & 0.0 & Escalate High to Urgent & 23.8 & Editorial Sprint & 69.6 & Q3 Feature Prep & 96.2 & Performance Backlog Push & 95.7 \\
    \bottomrule
  \end{tabular}}
  \caption{Five lowest-scoring sessions per model ($s^{overall}$ averaged across two runs).}
  \label{tab:worst-sessions}
\end{table*}